\documentclass[twocolumn]{aastex631}

\usepackage{hyperref}
\hypersetup{
    colorlinks=true,
    linkcolor=blue,
    filecolor=cyan,      
    urlcolor=magenta,
    citecolor=blue,
}

\chardef\us=`\_

\submitjournal{ApJL}

\begin{document}

\title{The Closest View of a Fast Coronal Mass Ejection: How Faulty Assumptions near Perihelion Lead to Unrealistic Interpretations of PSP/WISPR Observations}

\correspondingauthor{Ritesh Patel}
\email{ritesh.patel@swri.org}

\author[0000-0001-8504-2725]{Ritesh Patel}
\affiliation{Southwest Research Institute, 1050 Walnut Street, Suite 300, Boulder, CO 80302, USA}

\author[0000-0002-0631-2393]{Matthew J. West}
\affiliation{Southwest Research Institute, 1050 Walnut Street, Suite 300, Boulder, CO 80302, USA}

\author[0000-0002-0494-2025]{Daniel B. Seaton}
\affiliation{Southwest Research Institute, 1050 Walnut Street, Suite 300, Boulder, CO 80302, USA}

\author[0000-0003-1377-6353]{Phillip Hess}
\affiliation{U.S. Naval Research Laboratory, Washington, DC, USA}

\author[0000-0001-6692-9187]{Tatiana Niembro}
\affiliation{Center for Astrophysics $|$ Harvard \& Smithsonian, Cambridge, MA 02138, USA}

\author[0000-0002-6903-6832]{Katharine K. Reeves}
\affiliation{Center for Astrophysics $|$ Harvard \& Smithsonian, Cambridge, MA 02138, USA}



\begin{abstract}
We report on the closest view of a coronal mass ejection observed by the Parker Solar Probe (PSP)/Wide-field Imager for {Parker} Solar PRobe (WISPR) instrument on September 05, 2022, when PSP was traversing from a distance of 15.3~to~13.5~R$_\odot$ from the Sun. The CME leading edge and an arc-shaped {\emph{concave-up} structure near the core} was tracked in WISPR~field of view using the polar coordinate system, for the first time. Using the impact distance on Thomson surface, we measured average speeds of CME leading edge and concave-up structure as $\approx$2500~$\pm$~270\,km\,s$^{-1}$ and $\approx$400~$\pm$~70\,km\,s$^{-1}$ with a deceleration of $\approx$20~m~s$^{-2}$ for the later. {The use of the plane-of-sky approach yielded an unrealistic speed of more than three times of this estimate.}
We also used single viewpoint STEREO/COR-2A images to fit the Graduated Cylindrical Shell (GCS) model to the CME while incorporating the source region location from EUI of Solar Orbiter and estimated a 3D speed of $\approx$2700\,km\,s$^{-1}$.
We conclude that this CME exhibits the highest speed during the ascending phase of solar cycle 25. This places it in the category of extreme speed CMEs, which account for only 0.15\% of all CMEs listed in the CDAW CME catalog.


\end{abstract}

\keywords{Sun: corona --- Sun: coronal mass ejections --- Magnetic reconnection  --- Sun: coronal transients }


\section{Introduction} 
\label{sec:intro}

Coronal mass ejections (CMEs) span a wide range of speeds from a few tens to a few thousands of km\,s$^{-1}$ \citep{Webb2012LRSP}. Their collective average speeds vary from about 300\,km\,s$^{-1}$ during solar minimum to $\approx$500\,km\,s$^{-1}$ during solar maximum activity \citep{Zhang2021PEPS}. {It has been known that fast CMEs (i.e., those with speeds $>1000$\,km\,s$^{-1}$) experience the majority of their acceleration at heights near the Sun before being taken over by the ambient solar wind, while slower CMEs are mostly governed by the aerodynamic drag of the medium over their entire evolution \citep[][]{Sheeley1999, Gopalswamy2000, vrsnak2004A&A, Sachdeva2015ApJ, West2023}.}

A fraction of CMEs are known to reach extreme speeds of over 2000\,km\,s$^{-1}$ in the heliosphere \citep{Gopalswamy2016GSL}. {Only a few CMEs, such as the Halloween storm of October 28, 2003, are known to have extreme speeds and have resulted in extreme space weather consequences leading to solar energetic particle (SEP) events, and relatively strong shocks \citep[][and references therein.]{Gopalswamy2017}.} A possible upper limit on CME speeds of $\approx$4000\,km\,s$^{-1}$ has been established, based on the available free energy in source regions \citep{Gopalswamy2010SunGe}. The most recently observed CME with such extreme characteristic occurred on 2017~September~10, reaching a speed of $\approx$3000\,km\,s$^{-1}$ at heights beyond 20~R$_\odot$ \citep{Gopalswamy2018ApJ}, gaining most of that speed inside of 2~R$_\odot$ \citep{Seaton_Darnel2018}. On the other hand, CMEs in the heliosphere are also found to have speeds in the range of 100--2000\,km\,s$^{-1}$, with majority between 300--400\,km\,s$^{-1}$ \citep{Barnes2019SoPh}.

{The Large Angle Spectrometric Coronagraph \citep[LASCO;][]{Brueckner95} on the \textit{Solar and Heliospheric Observatory} (SOHO) has provided the longest synoptic observations of the corona, enhancing our understanding of coronal dynamics.} These observations were aided by stereoscopic analysis with the launch of telescope suites of the Sun Earth Connection Coronal and Heliospheric Imager \citep[SECCHI;][]{Howard2008SSRv} onboard the two \textit{Solar Terrestrial Relations Observatory} (STEREO) spacecraft. More recently launched instruments, including the Wide-Field Imager for Solar PRobe \citep[WISPR;][]{Vourlidas2016SSRv} on the \textit{Parker Solar Probe} \citep[PSP;][]{Fox2016SSRv}, and the Extreme Ultraviolet Imager \cite[EUI;][]{Rochus2020A&A} and \citep[Metis;][]{Antonucci2020A&A} and Solar Orbiter Heliospheric Imager \cite[SoloHI;][]{Howard2020} onboard Solar Orbiter \citep[SolO;][]{SolO2021A&A}, are providing new observations from different viewpoints that have already significantly enhanced our understanding of this extremely energetic phenomenon.

One of the challenges in using WISPR data to characterize transients like CMEs close to the Sun is the rapidly changing field of view (FOV) with the changing position of PSP. {Unlike the Heliospheric Imagers (HI) onboard the STEREO spacecraft, which observe the heliosphere from $\sim$1~au, the FOV changes for the WISPR telescopes. The WISPR combined FOV from 13.5$^\circ$ to 108$^\circ$ elongation corresponds to a varying heliocentric distance (e.g. see Figure 1 of \citet{Stenborg2022ApJ} for encounter 9). This also results in time varying spatial resolution in WISPR data during the complete period of observation \citep{Vourlidas2016SSRv}, hence the methods used for imagers at 1~au cannot be used directly for analysis.} To assess the effect of rapid motion along the highly elliptical orbit on tracking density features, \citet{Liewer2019} used synthetic white-light observations for streamer-like structures and radially outward-moving features to provide a basic framework for kinematic analysis. During the first day of the first PSP encounter on November 01, 2018, the first CME observed by WISPR was a slow-speed streamer blowout-type event \citep{Howard2019Natur, Hess2020}. Later, the first-ever 3D morphology of a CME {using WISPR data} was studied by combining the observations of the first encounter from WISPR with STEREO/SECCHI observations during this time period \citep{Wood2020}. 

\citet{Nindos2021} introduced methods based on the conventional elongation vs time J-map, with an extension to physical distance units where the elongation angles are converted to impact distance on the Thomson sphere to generate height-time map (R-map) to track various outflows in WISPR images. Recently, new methods have been  developed for tracking CME~3D~trajectories, such as using curve-fitting techniques \citep{Liewer2020}, and tracking transients based on vector calculations with WISPR and STEREO/HI observations \citep{Braga2021}.
WISPR also has provided an opportunity to study the internal structure of a few CMEs \citep{Wood2021, Liewer2021}.
\citet{Braga2022} identified the deformation of a CME at 0.1~AU after the self-similar expansion phase of its evolution ended. This deformation ultimately led to an error in arrival time estimation of the CME at 0.5~au.
At least 10~CMEs were observed by WISPR during encounter~10, the perihelion to reach $\sim$13.29 R$_\odot$ in November~2021, which was at the time the most CMEs observed during any PSP~encounter when compared with the previous ones \citep{Howard2022}. 

During encounter~13 from 1~to~11~September~2022, when PSP was cruising from 15.5~to~13.5~R$_\odot$, just prior to perihelion, a large and energetic CME, observed in WISPR~images, ultimately fully engulfed the PSP spacecraft. In this Letter, we present the kinematics of this CME using the WISPR~data-set from the encounter, along with other available white-light coronagraph imagery. We present the details of the observations in Section~\ref{sec:obs} and the analysis and results obtained using different methods for kinematic analysis in Section~\ref{sec:results}. We conclude and discuss the possibilities for future studies for this energetic CME in Section~\ref{sec:conclusion}.

\section{Observations}
\label{sec:obs}
\begin{figure*}
    \centering
    \includegraphics[width=0.97\linewidth]{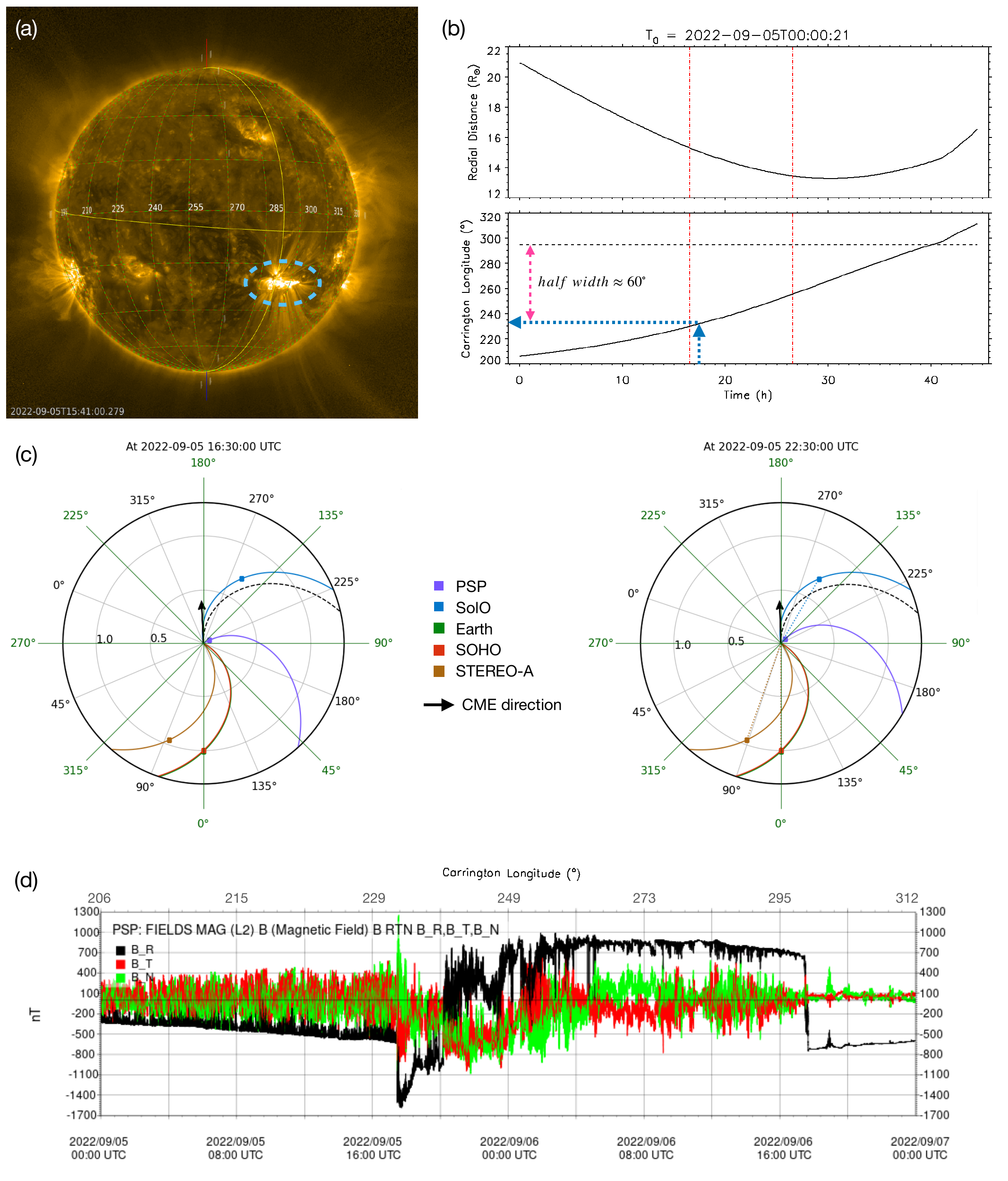}
    \caption{(a) SolO/EUI context image on 2022~September~05, where the source region for the eruption is highlighted by the blue dashed ellipse. (b) The top and bottom panels show the variation of radial distance and Carrington longitude of the PSP~spacecraft. The red dashed vertical lines indicate the time period during which the CME and its concave structure is tracked. (c) Positions of remote-sensing instruments at two instances when the CME is observed in WISPR FOV at 16:30 and 22:30 UT. (d) In-situ magnetic field measurement from PSP/FIELDS instrument on September~05~and~06.}
    \label{fig:context}
\end{figure*}











We observed the CME within a few hours of the so-called ``Solar Snake\footnote{\url{https://www.esa.int/ESA_Multimedia/Videos/2022/11/Solar_snake_spotted_slithering_across_Sun_s_surface}}" motion of plasma within a filament in SolO/EUI images on 2022-09-05. Using EUI data we confirmed that the CME erupted at $\approx$16:00\,UT, from an active region located around a longitude and latitude of $\approx$--171$^\circ$ and $\approx$--27$^\circ$ (Carrington longitude $\approx$295$^\circ$) respectively (Figure~\ref{fig:context}a). The trajectory of PSP, in the form of radial distance and Carrington longitude, appears in Figure~\ref{fig:context}b. The CME entered the WISPR-I field of view (FOV) at $\approx$16:30~UT when the spacecraft was at a distance of 15.3~R$_\odot$. A concave-up structure was also identified within the CME in the WISPR combined FOVs near the perihelion marking the closest view from 13.5~R$_\odot$. Even though PSP got as close as 13.28~R$_\odot$, the CME core faded and merged with the background signal by 13.5~R$_\odot$. A movie with the combined FOVs of WISPR during the encounter~13 can be found at \url{https://wispr.nrl.navy.mil/encounter13-summary}. Using the Aditya-L1 orbit tool\footnote{\url{https://al1ssc.aries.res.in/orbit-tool}}, we plotted the location of all spacecraft with imaging instruments (Figure~\ref{fig:context}c) for the time when CME first enters the WISPR FOV. 
The CME appeared as full halo in the STEREO coronagraphs, while only a couple of frames are available from LASCO due to a data outage. In-situ measurement of the magnetic field by the PSP/FIELDS instrument \citep{PSPFIELDS2016SSRv} suggests that the PSP made the first contact with the CME at $\approx$17:35 UT (Figure~\ref{fig:context}d). The same time is also marked in the panel (b) by blue arrow to highlight the location of PSP at that time.

\section{Analysis and Results}
\label{sec:results}

\begin{figure*}
    \centering
    \includegraphics[width=1.05\linewidth]{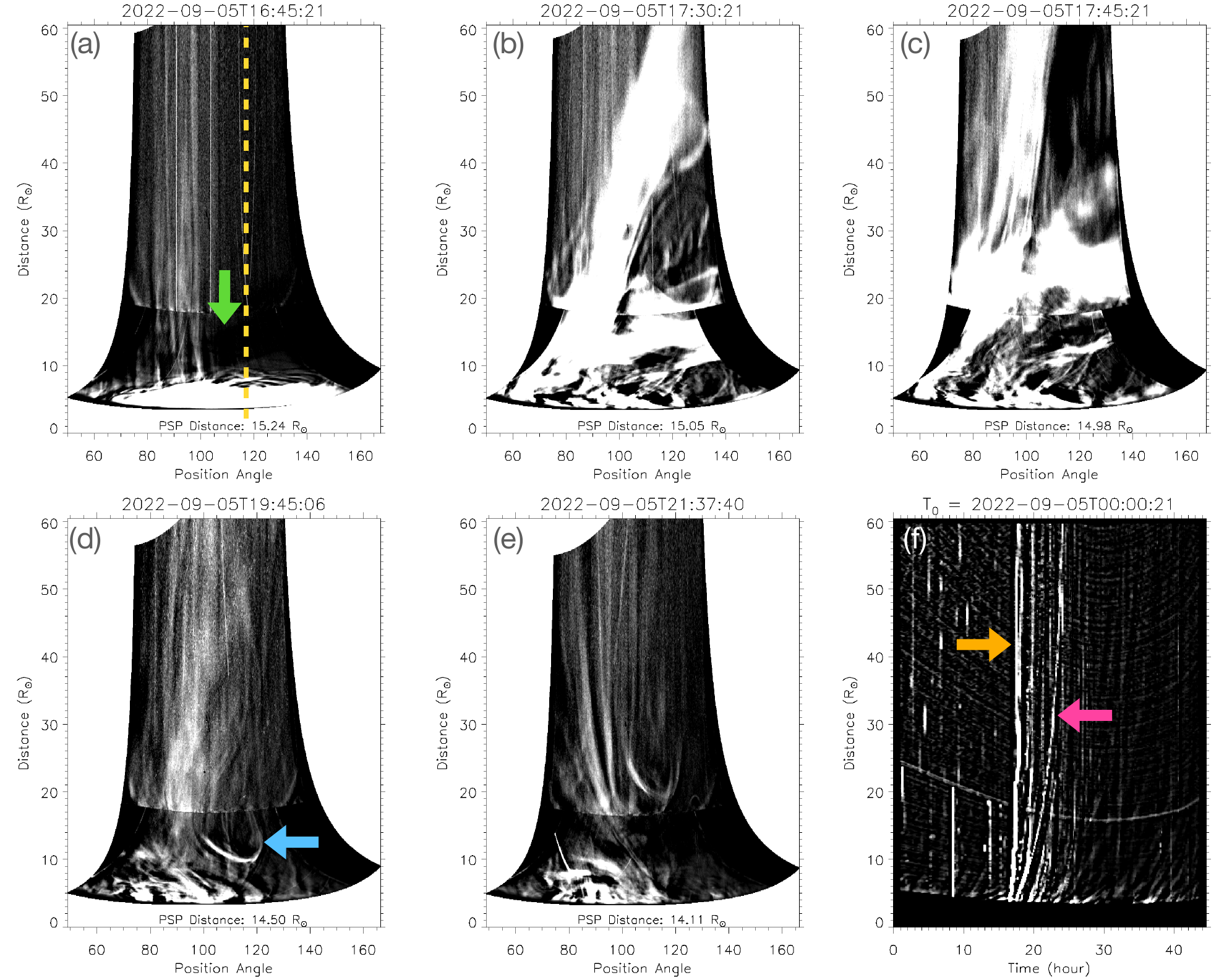}
    \caption{(a)-(e) Evolution of the CME in the WISPR FOV based on plane of sky projection in polar coordinates, where the concave-up structure associated with the CME core can be identified in (d) and (e). {The green arrow in (a) points to a dark fuzzy structure corresponding to CME shock candidate.} The blue arrow in (d) indicates the concave-up structure of CME core. (f) shows the height-time plot generated along the yellow dashed line marked in (a). The orange and magenta arrows indicate the ridges corresponding to the CME leading edge and concave-up structure, respectively. {The animation begins on 2022~September~05 at 12:00 UT and ends at 06:00 UT of the following day. The real-time duration of the animation is 12 s.} \\ An animation is available along with this figure showing the CME evolution in the POS~projection.}
    \label{fig:pos}
\end{figure*}

For this study, we used Level-3 WISPR~data, which have the F-corona background removed from calibrated Level-2 images and are in units of mean solar brightness \citep{Hess2021}. These images have a cadence of 15~minutes until 19:00~UT on 2022~September~5, after which the cadence was increased to 7.5~minutes during the period near perihelion. 
To suppress the background star-field, we processed the data with a sigma filter, which works in the following way: first, we compute the mean and standard deviation of intensities in the neighborhood of a given pixel defined by the choice of kernel~size (15$\times$15 pixels$^2$ in our analysis). If the central pixel intensity exceeds a set threshold, {based on neighbourhood pixels} (mean\,+\,3\,$\times$\,standard deviation), the intensity of the central pixel is replaced by the neighbourhood mean. This process is applied iteratively 20~times, until most of the background stars are removed. A benefit of this filter method over other approaches, like median filtering, is that it strongly suppresses significant outliers (like stars) while leaving the bulk of the image untouched. Because the CME~brightness varies only over relatively large spatial scales, while the stars are essentially point outliers, this technique is an especially helpful choice for these data.

\begin{figure*}
    \centering
    \includegraphics[width=0.97\linewidth]{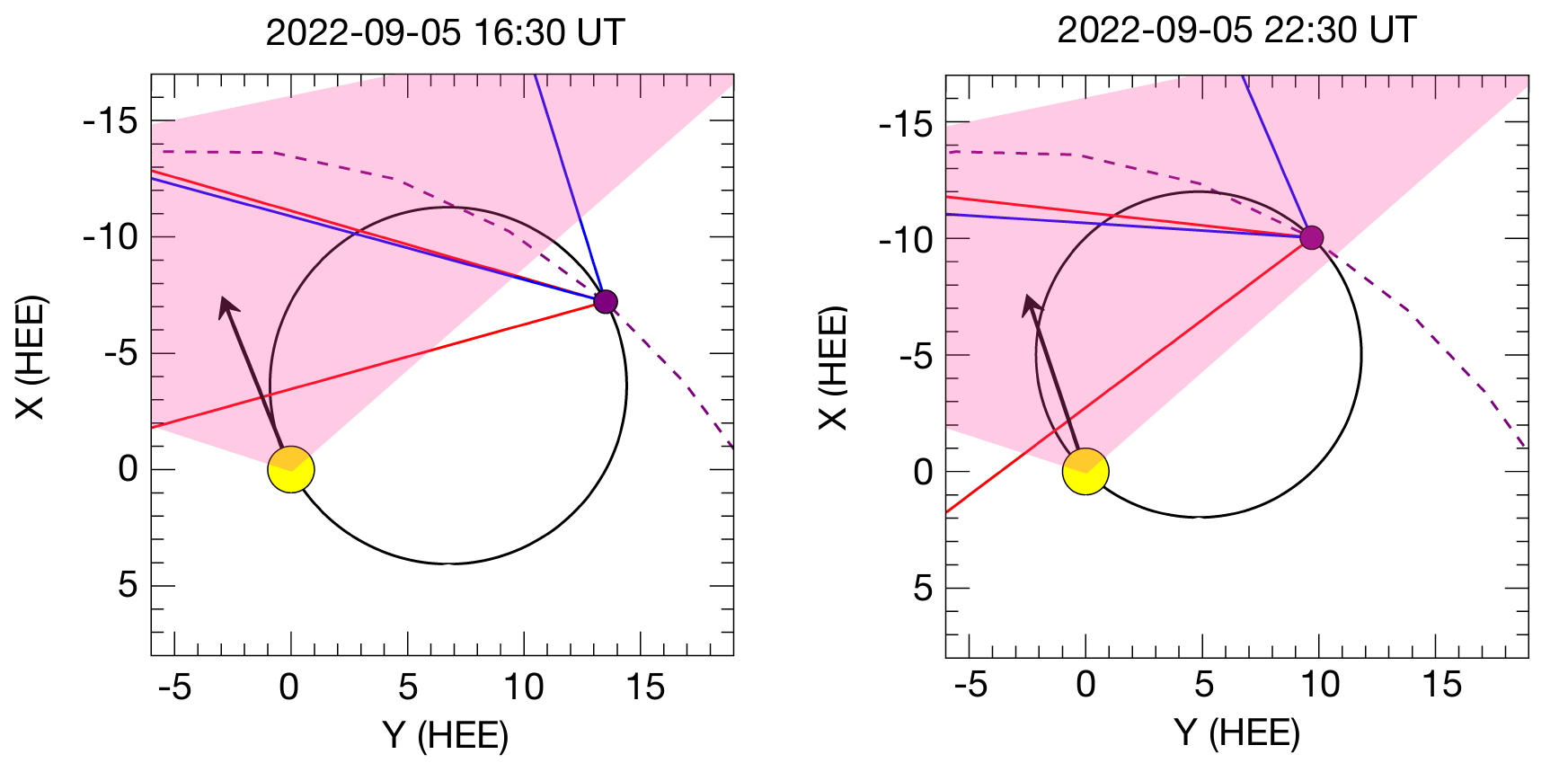}
    \caption{Close up view of PSP orbit at 16:30 UT and 22:30 UT respectively in Heliocentric Earth Ecliptic (HEE) coordinates where two axes are in R$_\odot$. The black circle represents Thomson surface while cones in red and blue outlines WISPER-I and O FOVs respectively. Magenta colour filled cone gives an extent of the CME width with black arrow indicating the direction of CME propagation based on source region information.}
    \label{fig:thomson}
\end{figure*}

The combined WISPR-I and WISPR-O data are initially projected in the Helioprojective Cartesian (HPC) system, where the columns and rows in images correspond to the elongation angle and latitude, respectively, in the spacecraft frame of reference. We converted the HPC projection to the Helioprojective Radial (HPR) system\footnote{See \citet{Thompson2006} for a complete discussion of different solar imaging coordinate systems and their properties}, using the same approach as described in \citet{Pant2016} for CME characterization in STEREO/HI-1 images. {The conventional method for analyzing CME kinematics based on plane of sky (POS) measurements has even been used with heliospheric imagers such as STEREO/HI-1. \citet{Liewer2019} demonstrated the challenges of this approach based on elongation maps that use synthetic WISPR datasets.
We use this CME as a case study first to illustrate the application of methodology similar to CACTusHI by \citeauthor{Pant2016} which was implemented for STEREO/HI-1.} During the change of coordinates, we transformed elongation angles to radial distances from the Sun in the POS. We subsequently converted this to polar projection with position angle measured in the clockwise direction from solar north, which we display along the horizontal axis, and radial distance from the Sun in vertical axis. Figure~\ref{fig:pos}a--e shows the POS-projected WISPR images in polar coordinates. An advantage of using a polar coordinate system is that height-time maps can be generated at any position angle and can be used to track radially outward moving structures. {It should be noted that this CME is associated with a shock that appears as a faint dark fuzzy structure indicated by the green arrow in Figure~\ref{fig:pos}a. The evolution of the this structure can be identified ahead of the CME leading edge in multiple frames in the corresponding animation from 16:00 to 17:00 UT.}

\begin{figure*}
    \centering
    \includegraphics[width=1.02\linewidth]{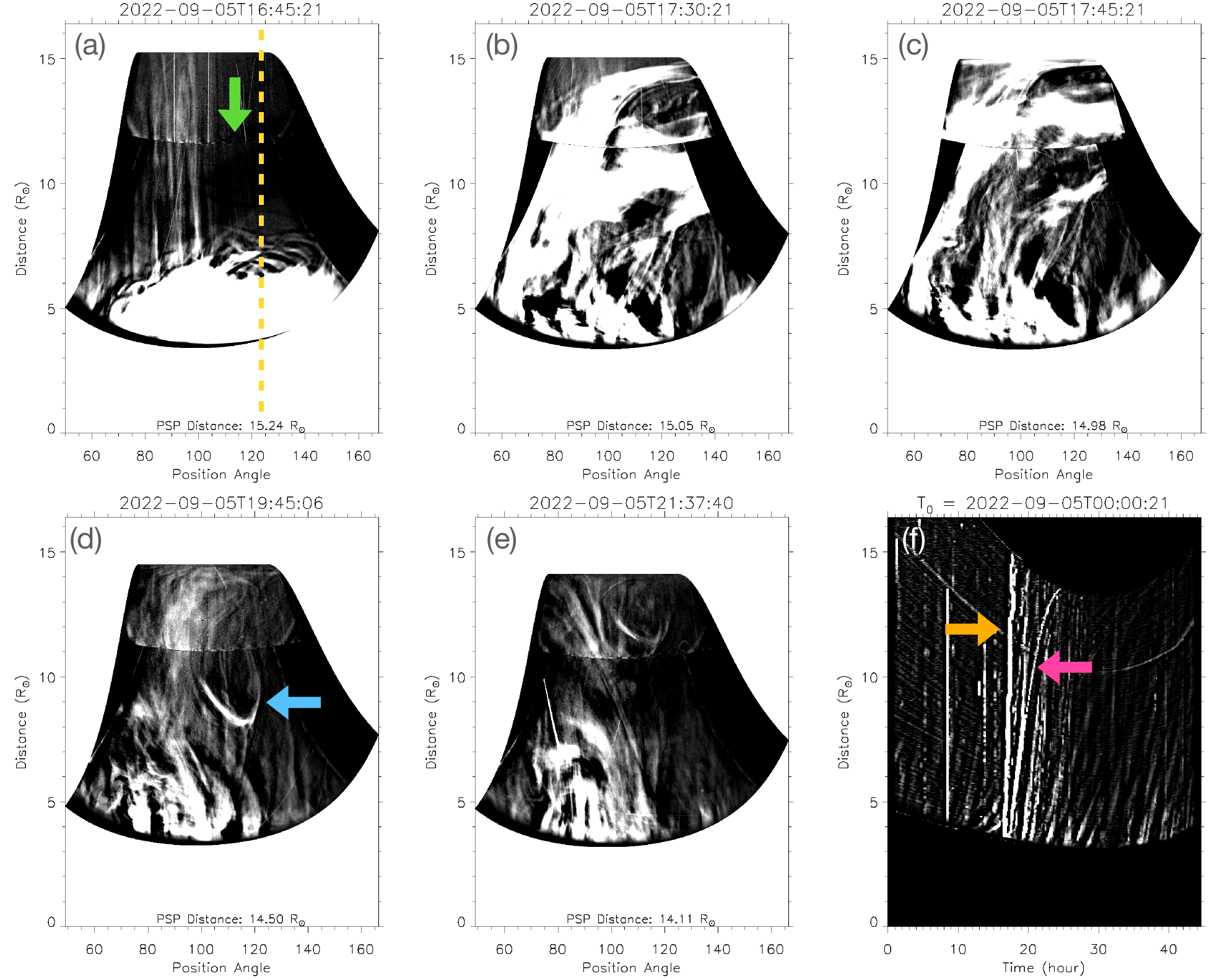}
    \caption{(a)-(e) Evolution of the CME in the WISPR FOV in polar projection based on the impact distance on the Thomson sphere. All  arrows indicate the same structures as in Figure \ref{fig:pos}.
    {The animation begins on 2022~September~05 at 12:00 UT and ends at 08:37 UT of the following day. The real-time duration of the animation is 19 s.} \\ An animation is available along with this figure showing the CME evolution in polar projection based on impact distance on the Thomson sphere.}
    \label{fig:impact}
\end{figure*}

We then generated the height-time plot at position angle of 116$^\circ$, near the center both of the leading edge and the center of the concave-up structure. As the imaging cadence changes when the CME is still in the WISPR FOV, we interpolated the height-time plots to obtain a uniform cadence of 7.5~minutes. The height-time plot is shown in Figure~\ref{fig:pos}f, where the first steep ridge (indicated by the yellow arrow) corresponds to the leading edge of the CME, whereas the curved ridge (magenta arrow) shows the evolution of the concave-up structure. {The ridge corresponding to the bright leading edge dominates the intensity scaling in R-maps, helping to distinguish the outward-moving CME structures from the shock.} We performed linear and parabolic fitting to these two ridges with both a manual and an automated method, the CMEs Identification in Inner Solar Corona \citep[CIISCO;][]{Patel2021,patel2022thesis}, to derive the kinematics. With these methods we found that the leading edge and concave-up structure propagated with an average speed of $\approx7800\pm970$\,km\,s$^{-1}$ and $\approx2000~\pm120$\,km\,s$^{-1}$, respectively. Our analysis shows that the concave-up structure had an acceleration of $\approx$130\,m\,s$^{-2}$. The leading edge speed in particular appears to be unreasonably fast, indicating that this simplistic approach to measuring the speed is inappropriate for the situation {as indicated by \citet{Liewer2019}}. Specifically, these values are measured in the plane of sky, and hence suffer from projection effects from the spacecraft viewpoint and rapidly changing perspective with respect to the CME itself as it progresses through the solar encounter.

\begin{figure*}
    \centering
    \includegraphics[width=1\linewidth]{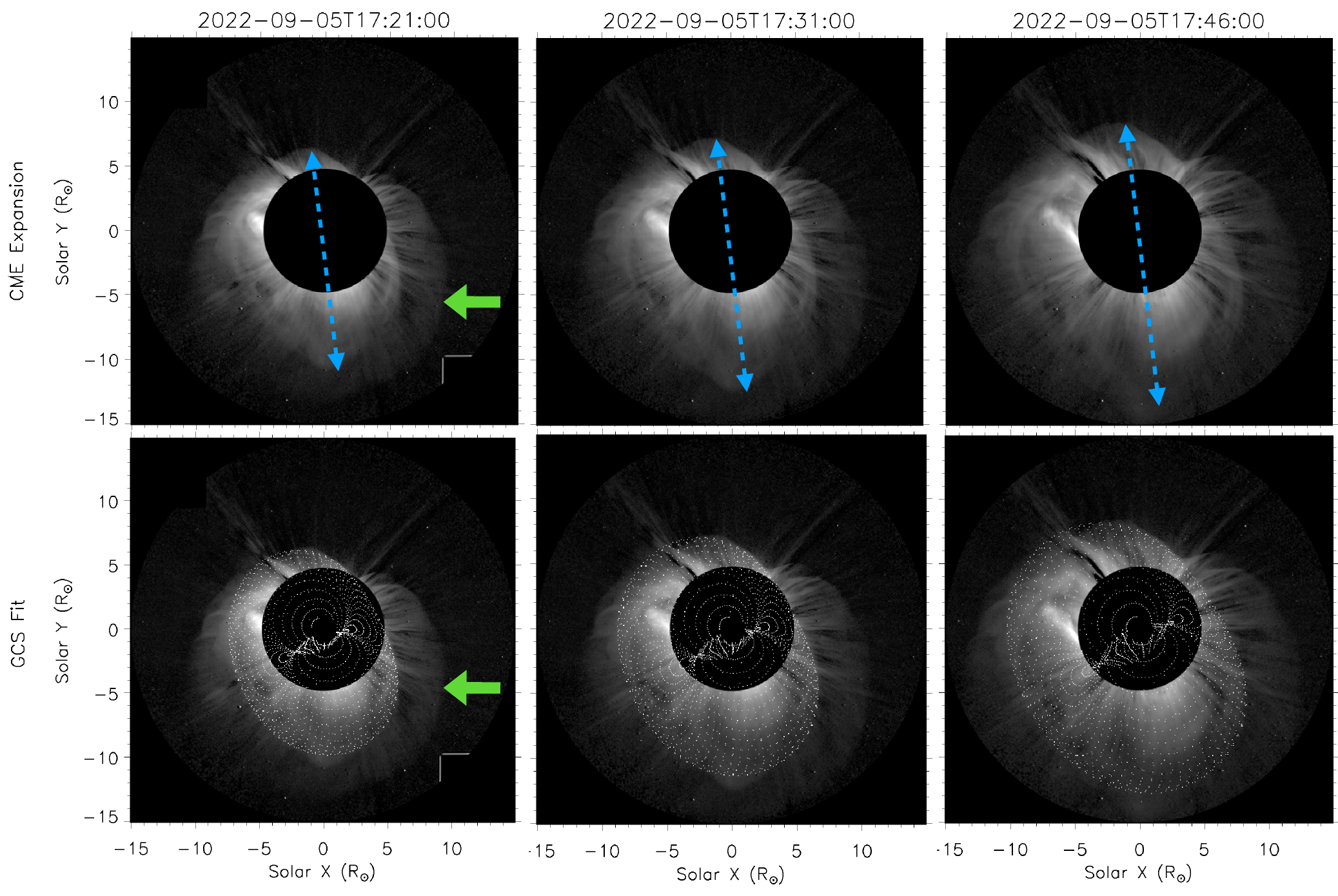}
    \caption{{\it Top:} Snapshots of CME observations from STEREO-A/COR-2 at three times. The blue dashed double-head arrow represents the extent of lateral expansion of the CME, similar to the methodology used by \citet{Michalek2009SoPh}. {\it Bottom:} The GCS fit at the same times shown with a white colored wire-frame of the fit superimposed on the COR-2A images. {The green arrow points to the shock in both the panels.}}
    \label{fig:gcs}
\end{figure*}

{A more reasonable approach to study such a close encounter of PSP with the CME is based on the principle of Thomson scattering.}
It is well known that in white-light observations the Thomson scattering strength is maximized when the line of sight from observer is perpendicular to the location of scattering electron. All such points lie on the surface of a sphere, known as Thomson sphere, the diameter of which is the distance between observer and source (Sun). The perpendicular distance from the Sun center to Thomson sphere is denoted by the impact distance to Thomson surface (represented by parameter `d' in Figure~1 of \citet{Vourlidas2006}) which is related to the elongation angle and distance between observer and the Sun center.  {A zoomed in view of the PSP orbit is shown in Figure~\ref{fig:thomson} where the Thomson sphere is represented by a black circle, and the magenta filled cones indicates the extent of the CME. The black arrows indicate the direction of propagation. The plots show the change in the position of PSP and hence the Thomson sphere for the two observation time frames, when the CME first entered WISPR FOV at 16:30 UT, and secondly when the concave-up structure was near the outer edge of WISPR FOV at 22:30 UT.} 
\citet{Nindos2021} have pointed out that impact distance on the Thomson surface, {based on the ``Point P'' method} illustrated by \citeauthor{Vourlidas2006} and \citet{Mishra2014ApJ}, provides a better way to estimate the kinematics of moving structures in wide FOV images. We therefore converted the WISPR images from the HPC system to HPR, using the impact distance instead of POS approach.

We then transformed these data into to the polar coordinate system, following the same approach as above. The projected images are shown in Figure~\ref{fig:impact}a--e. Figure~\ref{fig:impact}f shows the height-time plot corresponding to position angle of 116$^\circ$. We note that this height-time plot is equivalent to the R-map introduced in \citet{Nindos2021}. 
For our analysis, we identified the ridges corresponding to both the leading edge (yellow arrow) and the concave-up structure (magenta arrow). {It should be noted that the CME is an extended structure and this approach assumes it is located on the Thomson sphere where the scattering efficiency is maximized. The CME leading edge enters WISPR FOV around 16:30 UT and exits after 1.5 hours. During this period PSP traverses from 15.3 to 14.97 R$_\odot$ resulting in a small shift in Thomson sphere.}
Using the impact distance approach, we found that the leading edge has a speed of $\approx2500\pm270$\,km\,s$^{-1}$ {with no signature of any acceleration}. In our new analysis, the profile of the concave-up structure that appeared to be accelerating in the POS view now exhibits the opposite behaviour. We estimated an average speed of $\approx400\pm70$\,km\,s$^{-1}$ with a deceleration of  $\approx$20\,m\,s$^{-2}$ for this structure. {We would like to mention that the average kinematics of these structures remain within the uncertainty range when the position angle for measurement varied between 80$^\circ$ to 130$^\circ$ for the leading edge and across the width of the concave-up structure.}

As the CME appears to be full halo in COR-2A FOV, the STEREO-A observations provide an opportunity to measure its expansion speed. Using the points located at the extreme edges of the CME, we measured its lateral dimension {along the blue line in the top panel of Figure \ref{fig:gcs}} following the approach described by \citet{Michalek2009SoPh}. We tracked the lateral extent of the CME in subsequent frames, as shown in the top row of Figure~\ref{fig:gcs}, giving an expansion speed of $\approx$2400\,km\,s$^{-1}$. Using the empirical relation between the radial speed ($V_{\mathrm{rad}}$) and expansion speed ($V_{\mathrm{exp}}$) given by \citeauthor{Michalek2009SoPh}, $V_{\mathrm{rad}} = 1.17\,V_{\mathrm{exp}}$, the radial speed is estimated to be $\approx$2650\,km\,s$^{-1}$. On the other hand, \citet{Gopalswamy2009CEAB} established a alternative (and, arguably, contradictory) width dependence relation based on theoretical approach for $V_{\mathrm{rad}}$ and $V_{\mathrm{exp}}$, such that for CMEs of width greater than 120$^\circ$ the ratio between the two speeds is 0.82. Following this relation gives an average speed for the CME as $\approx$2000\,km\,s$^{-1}$. These two methods are limited respectively by the sample size considered and the generalised shape of CMEs taken as an ice-cream cone used for deriving the relationships. Recently, it has also been identified that these two speeds are dependent on the heliospheric conditions and hence on solar cycle \citep{Dagnew_2020}.

\begin{figure*}
    \centering
       \centerline{\hspace*{0.05\textwidth}
            \includegraphics[width=0.5\linewidth]{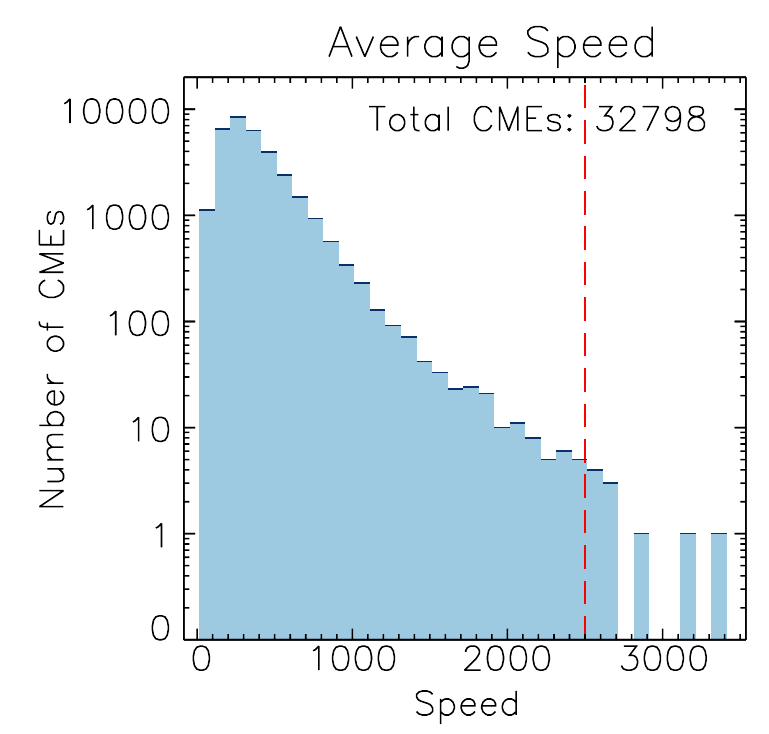}
            \includegraphics[width=0.465\linewidth]{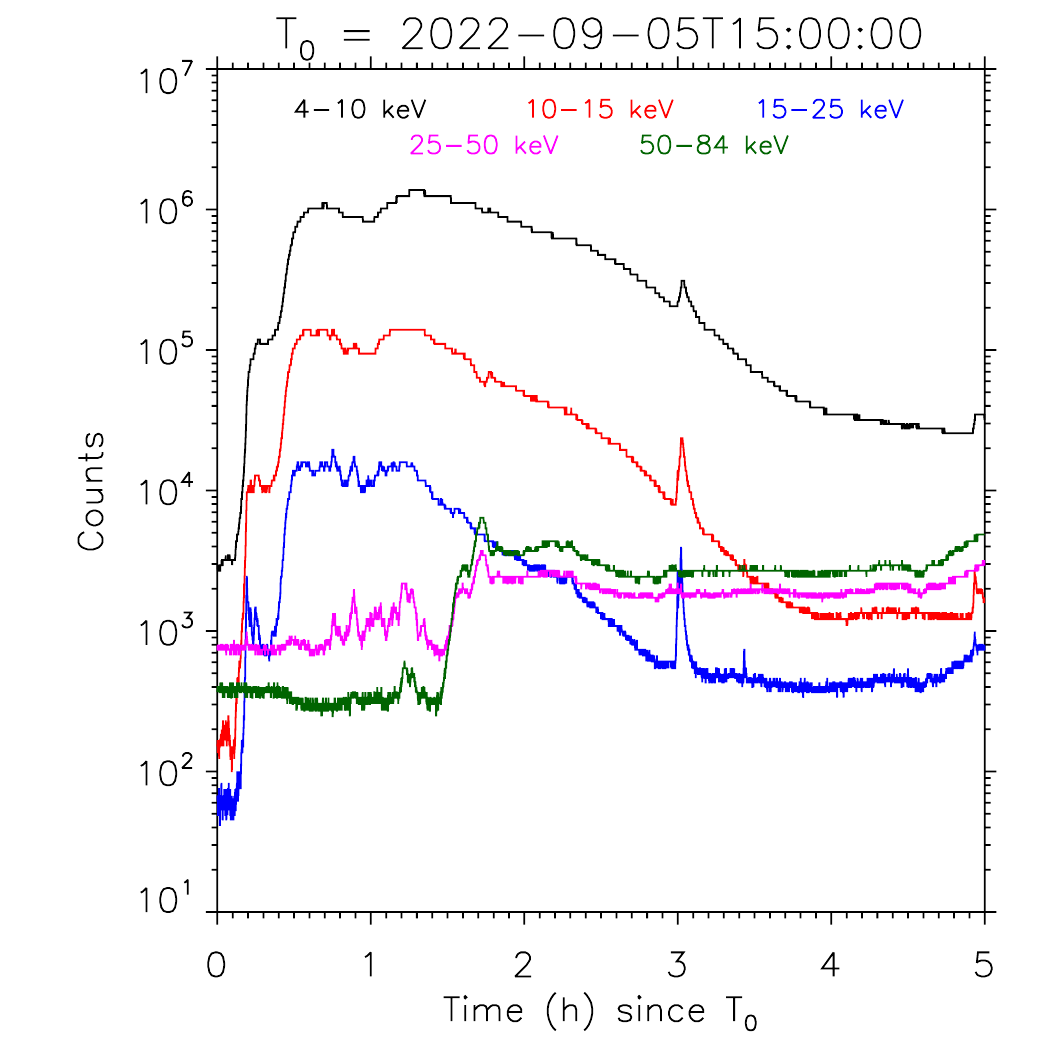}
              }
    
    \vspace{-.0385\textwidth} 
    \centerline{\large    
      \hspace{0.075\textwidth}  \color{black}{(a)}
      \hspace{0.44\textwidth}  \color{black}{(b)}
      \hfill}
      \vspace{0.0175\textwidth} 
    \caption{(a) Speed distribution of CMEs from the CDAW~catalog. The red dashed line is the approximate speed of the CME of 2022~September~5, (2500\,km\,s$^{-1}$). (b) SolO/STIX~X-ray~light-curve starting at 15:00~UT where the peak corresponds to the flare associated with the observed CME.}
    \label{fig:stix}
\end{figure*}

During this CME~observation, from 16:30~UT to 18:00~UT in STEREO-A/COR-2, the last image available from LASCO/C2 was at 16:46~UT, when the CME has just entered its FOV. The next frame was observed after 19:00~UT, making it essentially impossible to perform a 3D~reconstruction using using the Graduated Cylindrical Shell \citep[GCS;][]{Thernisien2006ApJ, Thernisien2011ApJS} model fit. The SolO/Metis instrument was also running with a low cadence of 2~hours during this period, resulting a temporal gap during our period of observation. 

The GCS model is optimized when multiple viewpoints can be included. While performing a GCS fit using only the COR-2A data increases the uncertainty significantly. 
{Nevertheless, SolO/EUI provides the location of the source region latitude and longitude. Assuming that such fast CMEs do not suffer from significant deflection, we constrain three of the six free parameters for a GCS fit of the COR-2A data: latitude, longitude and tilt angle, with relatively high fidelity. We also used the last available frame of LASCO/C2 at 16:48 UT along with the corresponding STEREO/COR-2A image for the initial GCS fit to the CME. Apart from these, we used the in-situ magnetic field measurement from PSP to get an estimate of the CME width complementing the initial fit based on these LASCO/C2 and STEREO/COR-2A co-temporal images. Figure~\ref{fig:context}(d) shows how the CME first hit the spacecraft, at around 17:35~UT. Considering the fact that the CME had a Carrington longitude of $\approx$295$^\circ$, shown by the horizontal dashed line in Figure~\ref{fig:context}(b), and at 17:35~UT PSP had a Carrington longitude of 232$^\circ$, we reasonably estimated that the CME could have a half-width of $\approx$63$^\circ$ and hence a full-width of $\approx$126$^\circ$. We use this information to constrain the half angle in the GCS fitting in subsequent STEREO/COR-2A frames, assuming that the CME has attained a constant width at the heights seen by WISPR \citep{Majumdar2020ApJ, Majumdar2022ApJ}. The aspect ratio parameter varied in the range of 0.45-0.6 which is coherent with majority of high speed CMEs observed \citep[e.g.,][]{Sachdeva2017SoPh}. 
Using all of these parameters, we fit the GCS model from the single vantage point of STEREO-A in consecutive frames. Snapshots for the fitted GCS model are shown in the bottom panel of Figure~\ref{fig:gcs}. Within the limits of the factors mentioned, we estimate a 3D~speed of $\approx$2700\,km\,s$^{-1}$, which, as expected, and is consistent with the projected speeds we measured using WISPR, albeit on the higher side.} 

The speed distribution of all the $>$32,000~CMEs from the Coordinated Data Analysis Workshop \citep[CDAW;][]{Yashiro04} catalog from January 1996 through December 2022 is shown in Figure~\ref{fig:stix}a. The red dashed line corresponds to our estimated CME velocity of 2500\,km\,s$^{-1}$. There are only 47~CMEs in the CDAW catalog with speeds greater than 2000\,km\,s$^{-1}$, and just 12 faster than 2500\,km\,s$^{-1}$ since the beginning of SoHO/LASCO observations, making the CME under consideration a rare and extremely energetic one. {We also conclude that the CME of 2022~September~5 is the first event with a speed of $\sim$2500 km s$^{-1}$ seen in solar cycle 25.}

As the CME kinematics are related to the flare at their source \citep{Zhang2001}, we plotted the X-ray light curve from the SolO/STIX instrument shown in Figure~\ref{fig:stix}b. It gives an impression that relatively low energy channels (represented in black, red and blue) show initial rise for the flare associated with this CME and by the time CME is seen in WISPR FOV at $\sim$16:30\,UT, they start to decrease in counts. The impulse at 18:00\,UT corresponds to flare from other region. The relation of the CME energetics with its source combining the SolO/EUI and STIX data needs a deeper analysis which is beyond the scope of the current study. 

\section{Conclusion and Discussion}
\label{sec:conclusion}
{In this Letter, we analyze the kinematics of the CME observed close-up by WISPR on 2022~September~05. While more analysis of the in-situ data is necessary to definitively understand the structures encountered by PSP, the compelling evidence in the images as well as the magnetic field data presented in Figure~\ref{fig:context} leads us to believe that this event represents the first instance where both imaging and in-situ measurements are available \citep[][submitted]{Romeo2023} for a highly energetic CME observed almost simultaneously in space and time from a single spacecraft located inside the corona.}


Importantly, we have illustrated the need for careful analysis of CMEs observed by WISPR when close to the Sun. We successfully demonstrated that the method based on the Thomson~Sphere impact distance is much more appropriate approach to estimate CME~speeds than the plane of sky (POS) measurements generally used for STEREO/HI \citep{Pant2016} analyses. Given the history of coronagraph observations of CMEs' POS velocities \citep[e.g.,][]{Yashiro04, Howard11}, such measurements would lead us to conclude it was a truly remarkable event with speed close to 8000\,km\,s$^{-1}$. Though the POS~method provides a simplistic approach for measuring the CME~kinematics, it breaks down for this event as this plane changes swiftly during PSP's closest approach to the Sun.

During this analysis, we developed {a pipeline} to convert WISPR~images to a polar coordinate system within its varying field of view. {The polar maps prove advantageous, as each column corresponds to a position angle and therefore height-time or R-maps can be generated at any position angle of interest. This provides a simple approach differing from the methods presented in \citet{Nindos2021} where angular sectors are used to generate elongation and R-maps.}
These maps and tools will be useful in the future for analyzing the flows, CMEs, plasma blobs and other radially moving structures identified by WISPR, during PSP~encounters. 

Although limited by data gaps, we demonstrated the value of multi-perspective, multi-messenger observations for developing deep understanding of events such as this one. We used the source region information from SolO/EUI and in-situ magnetic field measurements by PSP/FIELDS to (largely) constrain four of the six free parameters required by the GCS~model, and fit them using an additional vantage point with observations from STEREO-A/COR-2, confirming our direct measurements from the WISPR data.

Our estimates of a CME~speed of $\sim$2500\,km\,s$^{-1}$ using different methods implemented on WISPR and STEREO-A/COR-2 observations, suggest that this is the fastest CME observed by PSP since its launch. This CME is listed, with a speed of 2776\,km\,s$^{-1}$, in the CDAW~catalog based on just two frames, a remarkably accurate estimate given the severely limited LASCO~data for the event. 
Comparing this CME with the overall speed distribution from CDAW~CME~catalog, we found that this was one of the fastest CME observed since the dawn of coronagraph measurements.
The automated algorithms {Computer Aided CME Tracking \citep[CACTus;][]{Robbrecht04} and Solar Eruptive Events Detection System \citep[SEEDS;][]{Olmedo08}} could identify only a portion of the CME in LASCO and COR-2A images, and measured apparent speeds of 743\,km\,s$^{-1}$ and 893\,km\,s$^{-1}$, a significant underestimate.

It should be noted that during the first three years of the rising phase of all the recent solar cycles (SC) on record, there were just five CMEs with speeds greater than 2000\,km\,s$^{-1}$: two in SC23, and three in SC24. During SC25, prior to this event, only the CME of 2020~November~29 \citep{Nieves-Chinchilla2022} had a velocity (barely) over 2000\,km\,s$^{-1}$. With the CME speed estimated in this study, and also the CDAW~catalog value, this also becomes the only CME to exceed 2500\,km\,s$^{-1}$ during the rising phase of the three solar cycle observed. It is worth mentioning that of over 32,000~CMEs listed in the CDAW~catalog (up to December~2022), this CME is in extremely rare company, one of just 0.15\% of all CMEs having extreme speeds of exceeding 2000\,km\,s$^{-1}$.

{CMEs are known to show velocity dispersion across different sub-structures \citep[e.g.:][]{Colaninno2006ApJ, Ying2019ApJ, Li2021A&A}. {In some of these studies the CME leading edge was found to be moving with a speed more than 1000\,km\,s$^{-1}$ whereas sub-structures within it have speeds of the order of a few hundred \,km\,s$^{-1}$.  The CME presented in this study has been tracked both at its leading edge and at a concave-up sub-structure, hinting at the presence of a very dynamic event containing features moving at a range of speeds. This CME provides a detailed example of a complex structure, which can yield useful insights about the coherent -- or incoherent, as it may be -- nature of these events.} 
{It should be noted that the analysis presented in this study using Thomson sphere is based on the assumption that CME and its embedded structures all lie on this sphere. Since, PSP actually flies through the CME, some parts of this CME lie outside the sphere (Figure \ref{fig:thomson}). The estimation of speeds of the leading edge and the concave structure can be affected up to some extent.}
Nonetheless, the coupling between different sub-structures and their medium of propagation varies from ambient solar wind conditions for the leading edge to the CME internal plasma for internal parts will be an interesting study to be carried out in future.}

Energetics of such CMEs combined with the source region information are important to understand the role of surface activities for shaping the environment in the heliosphere. 
The decelerating nature of the concave-up structure at the core may suggest the reduction in the energy injected to propel the eruption. A comparison of its height-time profile (Figure~\ref{fig:impact}f) with the X-ray light curve from SolO/STIX (Figure~\ref{fig:stix}b) appears to hint at a decline in magnetic reconnection rate coincident with this deceleration, yielding decreased X-ray flux in relatively low energy channels and reducing the tether cutting required for the CME to escape. However, the EUI~data suggests the presence of an extended reconnection process similar to the observations presented in \citet{West2015ApJ}, who observed nearly two days of post-eruptive arcade growth following a CME in 2014. The availability of near-simultaneous imaging and in-situ measurements for this CME from PSP along with source region observations from remote sensing instruments onboard SolO, provide a unique opportunity to combine these observations for better understanding of the structures and phenomena within such rare extreme speed CME. An in-depth analysis is being carried out following this work to investigate the propulsion mechanism and underlying physical processes associated with this highly energetic CME.

\begin{acknowledgements}
{We thank the anonymous reviewers for their careful and insightful comments that have helped in improving the manuscript.} This work is funded by NASA Heliophysics System Observatory Connect program, grant number 80NSSC20K1283. We acknowledge Sam Van Kooten and Vaibhav Pant for fruitful discussions. Parker Solar Probe was designed, built, and is now operated by the Johns Hopkins Applied Physics Laboratory (APL) as part of NASA’s Living with a Star program (contract NNN06AA01C).
The WISPR instrument was designed, built, and is now operated by the US Naval Research Laboratory in collaboration with Johns Hopkins University APL, California Institute of Technology/Jet Propulsion Laboratory, University of Gottingen, Germany, Centre Spatiale de Liege, Belgium and University of Toulouse/Research Institute in Astrophysics and Planetology. Parker~Solar~Probe was designed, built, and is now operated by the Johns Hopkins Applied Physics Laboratory as part of NASA's Living with a Star (LWS) program (contract NNN06AA01C). WISPR is funded by NASA grant NNG11EK11I. The FIELDS experiment was designed and developed under NASA contract NNN06AA01C. We acknowledge the SOHO and STEREO teams to make the LASCO and SECCHI data available. The EUI instrument was built by CSL, IAS, MPS, MSSL/UCL, PMOD/WRC, ROB, LCF/IO with funding from the Belgian Federal Science Policy Office (BELSPO/PRODEX PEA 4000134088, 4000112292, 4000117262, and 4000134474); the Centre National d’Etudes Spatiales (CNES); the UK Space Agency (UKSA); the Bundesministerium für Wirtschaft und Energie (BMWi) through the Deutsches Zentrum für Luft- und Raumfahrt (DLR); and the Swiss Space Office (SSO). 
PH is also funded by the Office of Naval Research. TN was supported by the Parker Solar Probe project through the SAO/SWEAP subcontract 975569.
\end{acknowledgements}

\bibliography{reference}{}
\bibliographystyle{aasjournal}



\end{document}